# Bond-randomness-induced Néel order in weakly coupled antiferromagnetic spin chains

A. Joshi[1] and Kun Yang[1,2]

[1]*National High Magnetic Field Laboratory, Florida State University, Tallahassee, Florida 32310*
[2]*Department of Physics, Florida State University, Tallahassee, Florida 32306*

Quasi-one-dimensional antiferromagnetic (AF) quantum spin systems show a wide range of interesting phenomena such as the spin-Peierls transition and disorder driven long-range ordering. While there is no magnetic long-range order in strictly one-dimensional systems, in real systems some amount of interchain coupling is always present and AF long-range order may appear below a Néel ordering temperature $T_N$. We study the effect of bond randomness on Néel ordering in weakly coupled random AF $S=1/2$ chains, both with and without dimerization (or spin-Peierls order). We use the real space renormalization group method to tackle the intrachain couplings, and a mean-field approximation to treat the interchain coupling. We show that in the nondimerized chain, disorder (represented by bond randomness) enhances the Néel order parameter; in the dimerized chain which shows no magnetic ordering in the weak interchain coupling limit without randomness, disorder can actually lead to long range order. Thus disorder is shown to lead to, or enhance the tendency toward long range order, providing another example of the order-by-disorder phenomenon. We make a qualitative comparison of our results with the observed phenomenon of doping induced long-range ordering in quasi-one-dimensional spin systems such as $CuGeO_3$.

Quasi-one-dimensional quantum spin systems have been under intense theoretical and experimental investigation over the past few decades.[1] These systems show a wide range of interesting and unexpected phenomenon such as the spin-Peierls transition and disorder driven long-range ordering.[2–11] In strictly one-dimensional systems, there is no long-range magnetic order due to strong quantum and thermal fluctuations. However in real systems, such as $CuGeO_3$, $KCuF_3$, or $Sr_2CuO_3$ some amount of interchain coupling is always present.[3,12,13] In these cases magnetic long range order may appear below a Néel ordering temperature $T_N$. In this work we study the effect of disorder on long range ordering in weakly coupled spin chains. We show that disorder leads to, or enhances the tendency toward long-range order, providing another example of the order-by-disorder phenomenon that has received considerable attention in recent years.[14] The disorder driven long-range ordering seen here is purely quantum mechanical in origin, as opposed to being thermally driven.[14]

The specific example that we bear in mind in our study is the interesting phenomenon of doping driven long range ordering observed in the spin-Peierls material $CuGeO_3$. In this system the Cu ions form an effective one-dimensional (1D) antiferromagnetic (AF) spin-1/2 chain with weak interchain coupling. This is the first inorganic system to show the spin-Peierls transition.[3] Surprisingly, while the pure system has no magnetic order in the ground state, doping the system with a very small amount of impurities, such as Si for Ge (Refs. 4–10) or Zn, Ni, or Mg for Cu,[4,5,8,11] leads to the appearance of AF long range order (LRO) that may *coexist* with spin-Peierls order (or dimerization) below a Néel ordering temperature $T_N$. Increasing the doping level can drive the spin-Peierls transition temperature $T_{sP}$ to zero, while $T_N$ remains finite; eventually strong enough doping suppresses both spin-Peierls and Néel ordering in the system.

Here we study weakly coupled spin-1/2 AF chains with and without dimerization in presence of disorder, which are appropriate for the cases with and without spin-Peierls order respectively. The effect of doping in the system is represented by randomness in the intrachain couplings.[15] The approach involves treating the interchain coupling in a mean field approximation,[16,17] and solving the resulting effective 1D problem using the real space renormalization group approach (RSRG).[18,19] Using the asymptotically exact nature of RSRG in the limit of weak interchain coupling, within the mean-field approximation we are able to obtain analytical results in closed form in this limit, including the leading dependence of the Néel order parameter ($m$) and $T_N$ on the interchain coupling strength, and the $T$ dependence of $m$. We find that in the nondimerized chain, disorder enhances the tendency toward AF LRO. In the dimerized chain which shows no magnetic ordering in the pure case, disorder can actually lead to magnetic LRO. In real systems, on the other hand, interchain couplings may not be so weak. In this case we implement the RSRG and solve the mean-field equations numerically, and obtain results that show good qualitative agreement with behavior seen in experiments. In the rest of this paper we first discuss our model, the approximations used, and present our results; then we draw comparisons with the phenomenology of $CuGeO_3$, discuss the relation between our work and existing theoretical work on this subject, and possible extensions of our approach to other problems.

The chains are taken to be arranged parallel to one another, with $z$ nearest neighbors for each chain. The system is represented by the Heisenberg antiferromagnetic Hamiltonian with interchain coupling

$$H = \sum_{i,\vec{n}} J_{i\vec{n}}(\mathbf{S}_{i,\vec{n}} \cdot \mathbf{S}_{i+1,\vec{n}}) + J_\perp \sum_{i,\vec{n},\vec{\delta}} (\mathbf{S}_{i,\vec{n}} \cdot \mathbf{S}_{i,\vec{n}+\vec{\delta}}), \quad (1)$$

where $i$ is the site index along the chain, $\vec{n}$ is the chain index, and $\vec{\delta}$ is the index summed over the nearest neighbors. The intrachain coupling constants $J_{i\vec{n}}$ are site dependent and follow certain random distribution function, while the coupling between the chains $J_\perp$ is taken to be a constant. We consider $J_{i\vec{n}} > 0$ and $J_\perp > 0$; however, the conclusions remain valid for cases with $J_\perp < 0$. We treat the interchain coupling in the mean field approximation, thus its effect is approximated as a staggered field when there is long-range Néel order. We further assume the resulting staggered field to be uniform, thus we are effectively taking the limit of large coordination number (or large dimensionality) and averaging both the spatial variation of the field and the quantum fluctuations over the neighboring chains.

With these approximations the Hamiltonian $H$ [Eq. (1)] is transformed to an effective 1D Hamiltonian

$$H_1 = \sum_i J_i \mathbf{S}_i \cdot \mathbf{S}_{i+1} - h \sum_i (-1)^i S_i^z, \quad (2)$$

where the magnitude of the staggered field $h$ is determined by the self-consistency condition

$$h = z J_\perp |m|. \quad (3)$$

Here $m$ is the disorder-averaged staggered magnetization, which serves as the order parameter. The staggered magnetization at site $i$ is $m_i = (-1)^i \langle S_i \rangle$.

In order to solve the effective one-dimensional problem and obtain the order parameter $m$ we use the real space renormalization group (RSRG) method.[18,19] In this approach, the strongest bond in the distribution and its two weaker neighbors are progressively decimated and substituted by a much weaker effective bond. This is a perturbative procedure where the bonds neighboring the strongest bond are treated as perturbations. It has been shown[18,19] that for generic initial bond distributions without dimerization and in the absence of the staggered field, under RSRG the distribution of couplings very quickly flows to the so-called random singlet fixed point distribution, which is a broad power law distribution

$$P(J, \Omega_0) = (\alpha/\Omega_0)(J/\Omega_0)^{(-1+\alpha)} \theta(1 - J/\Omega_0), \quad (4)$$

with $\alpha$ being a positive exponent that *decreases* as energy scale is reduced; the perturbative RG becomes asymptotically exact in the low-energy limit.[18,19] $P(J,\Omega_0) dJ$ is the probability of finding a coupling between $J$ and $J+dJ$, $\Omega_0$ is the cutoff for the distribution, and $\alpha = 1/|\ln(\Omega_0)|$ in the limit $\Omega_0 \to 0$. The initial cutoff for the original distribution is used as the unit for both intrachain and interchain couplings and set to be 1, and the cutoff whereafter the distribution can effectively be treated as a power law is labeled $\Omega_c$. The fraction of undecimated spins at cutoff $\Omega_0$ ($\Omega_0 < \Omega_c$) is given by $N(\Omega_0) = N_c [\ln(\Omega_c)/\ln(\Omega_0)]^2$, where $N_c$ is the fraction of undecimated spins at $\Omega = \Omega_c$.

The RG process in the presence of a staggered field is more involved. In this case the field breaks the rotational symmetry of the system, thus Ising terms are generated in the renormalized Hamiltonian and the staggered field itself gets renormalized.[18] In the weak interchain coupling (and thus weak field) limit, however, we can show that these effects are very weak and therefore need not to be taken into account in our analysis in this limit; we do, however, include these effects when we address the more general problem numerically.

We first approach the problem from the weak interchain coupling limit. In this limit, the results obtained from the renormalization group approach are asymptotically exact and we can obtain analytical expressions for the dependence of the zero temperature order parameter ($m_c$) and the Néel ordering temperature $T_N$ on the interchain coupling, as well as the dependence of the order parameter on temperature $m(T)$. The case of the dimerized and nondimerized chains are treated separately.

We first consider the chain with no dimerization. As the RG is carried out, the spins are progressively eliminated and their contribution to the net staggered magnetization summed. At some stage of the RG the cutoff becomes comparable to the staggered field. At this point almost all the spins are fully polarized since most couplings in the power law distribution are much weaker than the staggered field and it is not necessary to proceed with the renormalization any further. For a given staggered field $h$, the contributions to the order parameter can be broken into three distinct parts. The first part is the contribution from initially decimated spins with $J_i > \Omega_c$. This contribution, which depends on the original coupling distribution and is nonuniversal, is of order $h/\Omega_c$ and negligible compared to other contributions (see later). The second contribution ($m_1$) arises from the decimated spins with coupling $h < J_i < \Omega_c$ and is given by

$$m_1 = chg(2h), \quad (5)$$

where $c = N_c \ln^2(\Omega_c)$, and $g(y) = \int_{x_1}^{x_2} (e^x/x^3) dx$ with $x_1 = |\ln(\Omega_c)|$ and $x_2 = |\ln(y)|$. The most significant contribution ($m_0$) to the order parameter results from the undecimated, almost fully polarized spins with coupling $J_i < h$. This contribution is given by

$$m_0 = c/[2 \ln^2(2h)]. \quad (6)$$

While there can arise some error in the estimate for $m$ resulting from the treatment of spins with $J_i \sim h$, we can show that in the weak coupling limit this error is small since the fraction of such spins is vanishingly small.

Combining Eqs. (5) and (6) with the self-consistency condition Eq. (3), we find the zero temperature order parameter ($m_c$) satisfies the equation

$$m_c = c\{1/[2 \ln^2(2h)] + hg(2h)\}$$
$$= c\{1/[2 \ln^2(2zJ_\perp m_c)] + zJ_\perp m_c g(2zJ_\perp m_c)\}. \quad (7)$$

The second term is subdominant to the first and negligible in the limit $J_\perp \to 0$; thus to leading order we obtain

$$m_c = c\{1/2 \ln^2(2zJ_\perp) + O[1/\ln^3(2zJ_\perp)]\}. \quad (8)$$

Thus, we find that the order parameter $m$ for the AF spin-1/2 chain with disorder shows a very singular dependence on the interchain coupling. By comparison, the uniform chain shows a much less singular power law dependence of the AF order parameter ($m \sim J_\perp^{1/2}$),[17] thus the Néel order parameter is much bigger in the random case in the weak coupling limit. This is because disorder leads to much larger density of low lying states in the energy spectrum and much more singular staggered susceptibility, thereby strongly enhances the tendency toward long-range magnetic order.

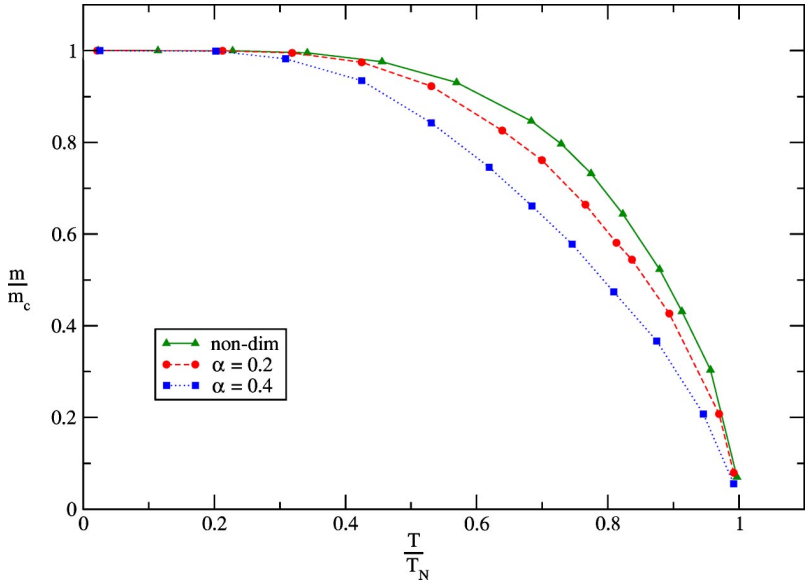

FIG. 1. Representative renormalized order parameter curves for the dimerized and nondimerized chains in the weak coupling limit. The triangles represent data for the nondimerized chain [generated using $zJ_\perp/J_0 = 0.0002$ in Eqs. (7) and (8), where $J_0$ is the cutoff for the initial intrachain coupling distribution; however, the choice of $zJ_\perp$ is not important to the results as long as $zJ_\perp \ll J_0$], circles for weak dimerization ($\alpha = 0.2$), and squares for strong dimerization ($\alpha = 0.4$). The curve for the nondimerized chain is universal, while that for the dimerized chains is governed by the single parameter $\alpha$.

For the case of finite $T$, we need to take into account the energy scale introduced by temperature. We consider two different regimes temperature regimes $T \gg h$ and $T \ll h$. For the case with $T \gg h$, the spins bound with couplings weaker than the temperature ($J_i < T$) are treated essentially as free spins, while those with $J_i > T$ are taken to form singlets. In the regime with $T \ll h$, the spins bound with $J_i > h$ are taken to form singlets, while those with $J_i < h$, which are almost fully polarized, are treated as free spins in high field $h$. The order parameter is governed by two distinct self consistency conditions in the two temperature regimes

$$m = \frac{c}{2}[\tanh(zJ_\perp m/2T)/\ln^2(4T)], \quad T \gg h, \quad (9)$$

$$m = \frac{c}{2}[\tanh(zJ_\perp m/2T)/\ln^2(2a_1 zJ_\perp m)], \quad T \ll h. \quad (10)$$

Here $a_1$ is an adjustable parameter that determines the energy scale at which we choose to stop the RG process when the field becomes comparable to the cutoff of the distribution. The low field limit ($h \ll T$) gives the dependence of the Néel ordering temperature on the interchain coupling

$$1 = czJ_\perp/[4T_N\ln^2(4T_N)], \quad (11)$$

while the low temperature limit gives back the leading dependence of the zero temperature order parameter. This leads to the universal relation between $T_N$ and $m_c$ that is valid in the weak interchain coupling limit

$$T_N = zJ_\perp m_c/2. \quad (12)$$

Using the two self-consistency conditions (7) and (8) we can obtain the finite temperature magnetization in the temperature regimes $T \gg h$ and $T \ll h$, respectively. However, in the region $T \sim h$ there is ambiguity about which of the two equations gives the correct magnetization. We resolve this issue by setting the adjustable parameter $a_1$ so that both equations lead to a consistent value of the order parameter in this temperature regime. Using this approach we obtain a universal order parameter curve for the random nondimerized chain in the weak coupling limit (Fig. 1).

We next consider the chain with dimerization. The random dimerized chain has a distribution of couplings with, say odd bond stronger than even bonds on an average. Due to preferential elimination of the odd bonds in the RG process, the distribution of even bonds becomes weaker faster than the odd bonds. At some value of the distribution cutoff $\Omega = \Omega_d$, the system goes into a random dimer phase, where it can be viewed as a collection of isolated spins pairs connected by odd bonds.[21] We consider the situation where the bond distribution reaches power law prior to the random dimer phase, which is the case for weak dimerization (as appropriate for $CuGeO_3$), and both cutoffs are much greater than the staggered field ($\Omega_c > \Omega_d \gg h$). In this case the distribution of the odd bonds in the low-energy limit takes the form of Eq. (4), with a fixed power-law exponent $\alpha$. The weaker the dimerization (or equivalently, the stronger the randomness for fixed dimerization), the smaller $\alpha$ is. For $0 < \alpha < 1$ the susceptibility (uniform or staggered) of the system *diverges* in the limit $T \to 0$; the system is in the Griffith phase in this case.[21]

Considering the system to be in the random dimer phase, and accounting for contributions as before we can obtain the results for the dimerized chain. The dependence of the zero temperature order parameter $m_c$ is given by

$$m_c = [c/2\ln^2(\Omega_d)]\left[\frac{1}{1-\alpha}\{(2zJ_\perp m_c/\Omega_d)^\alpha - \alpha(2zJ_\perp m_c/\Omega_d)\} + 2zJ_\perp m_c g(\Omega_d)\right]. \quad (13)$$

For $\alpha<1$ there is always a nontrivial ($m\neq 0$) solution, indicating an ordered ground state. Also the second and third terms are subdominant compared to the first term for small $J_\perp$ in this case. Thus, in this case the order parameter has power law dependence on the interchain coupling

$$m_c=[c/2\ln^2(\Omega_d)]\left[\frac{1}{1-\alpha}(2zJ_\perp m_c/\Omega_d)^\alpha\right]. \quad (14)$$

We saw earlier in the case of the nondimerized chain that disorder strongly enhances the tendency toward AF LRO. For the dimerized chain the effect of disorder is even more dramatic. The uniform dimerized chain has a ground state with a gap in the spectrum, and the system has no AF LRO for weak interchain couplings, due to the *vanishing* susceptibility at $T=0$. In the presence of sufficiently strong randomness in the bond distribution, however, the system is in a Griffiths phase with *divergent* susceptibility at $T=0$, and the coupled chains exhibit AF LRO at weak coupling; i.e., the AF LRO is induced by randomness in this limit.

The results for finite $T$ for dimerized chain are obtained using an analysis similar to the one used for the nondimerized chain and the order parameter is governed by two distinct self-consistency conditions

$$m=[c/2\ln^2(\Omega_d)][(4T/\Omega_d)^\alpha\tanh(zJ_\perp m/2T)], \quad T\gg h, \quad (15)$$

$$m=[c/2\ln^2(\Omega_d)]$$
$$\times\left[\frac{1}{1-\alpha}(2zJ_\perp m/\Omega_d)^\alpha\tanh(zJ_\perp m/2T)\right], \quad T\ll h. \quad (16)$$

The low field limit yields the dependence of the Néel temperature on the interchain coupling

$$T_N=\{[c/2\ln^2(\Omega_d)](4/\Omega_d)^\alpha(zJ_\perp/2)]\}^{1/(1-\alpha)}, \quad (17)$$

while the low temperature limit gives back the leading dependence of the order parameter. The dependences of the Néel temperature $T_N$ and the zero temperature order parameter $m_c$ leads to the relation

$$T_N=zJ_\perp m_c[(1-\alpha)^{1/(1-\alpha)}]/2. \quad (18)$$

The analysis above is similar to the nondimerized chain and gives us the analytical expressions that determine the order parameter in the two temperature regimes. However, for the dimerized chain the analysis can be simplified by the fact that in the random dimer phase the system can be considered as a collection of isolated spin pairs. This problem can be numerically solved exactly to obtain the order parameter at any given temperature. The advantage of this approach is that while it does not give closed form expression for determining the order parameter, the result is applicable in all temperature regimes; thus there is no ambiguity in obtaining order parameter for temperature regime $T\sim h$. A representative set of order parameter curves for the dimerized and nondimerized chains obtained in the weak coupling limit are shown in Fig. 1.

Just as in the undimerized case (12), the relation (18) for the dimerized chain predicts a linear dependence of $T_N$ on $m_c$, with a proportionality constant that depends on the interchain coupling and $\alpha$. It should be possible to experimentally verify both the predicted linear nature of the dependence and the value of the proportionality constant from this result. The value of $\alpha$ needed to determine the proportionality constant can be determined experimentally from the temperature dependence of the staggered magnetization by fitting the theoretical renormalized order parameter curve (which depend on $\alpha$) to the one obtained experimentally. Using the fitted value of $\alpha$ and the known interchain coupling, the proportionality constant can be calculated and matched with experiments. Another way to determine the exponent $\alpha$ experimentally is through the slope of the $m$-$T$ curve of Fig. 1 at $T=T_N$. Theoretically, this slope is given by the relation

$$\frac{dm^*}{dT^*}=\sqrt{3}[(1-\alpha)^{1/(1-\alpha)}]\{[2-(3-\alpha)(T^*)^{(1-\alpha)}]/$$
$$[1-(T^*)^{(1-\alpha)}]\}, \quad (19)$$

where $m^*=m/m_c$ and $T^*=T/T_N$. Thus from the measured values of $m_c$, $T_N$ and the slope we can solve for $\alpha$. The consistency of the different ways of obtaining $\alpha$ can also be used as a test of our theory. However, it should be noted that the result Eq. (18) is obtained in weak interchain coupling limit within the mean field approximation and hence may not be applicable to all systems. Experiments on Zn doped $CuGeO_3$ indeed observe a linear dependence of $T_N$ on $m_c$.[39] As will be discussed later, our approach is more appropriate for the case of Si doping. It would be very interesting to compare the results obtained here with experimental results for the case of Si doping in $CuGeO_3$ if such measurements are made.

The analytical approach allows us to obtain the dependences of the order parameter and the Néel temperature on the interchain coupling, as well as the dependence of the order parameter on temperature in the weak coupling limit. However, in real systems the interchain couplings may not be so weak. In order to address the more general problem involving for example a larger range of interchain coupling, we implement the RSRG and solve the mean-field equations numerically. In addition to addressing the problem of larger interchain coupling, this approach additionally allows us to include the effect of the renormalization of the staggered field and the Ising terms that are generated in the RG process in presence of a staggered field. In this method we start with a choice of the initial distribution of couplings and the interchain coupling strength. These choices determine the problem we are trying to address; for example, in the case of the dimerized chain, the even bonds are assigned a weaker distribution of couplings than the odd bonds. The RG process is then carried out by progressively decimating the strongest bond in the distribution. Within this approach, it is possible to study systems involving coupling distribution following a specific spatial pattern and explore the spatial variation of the resultant order parameter.

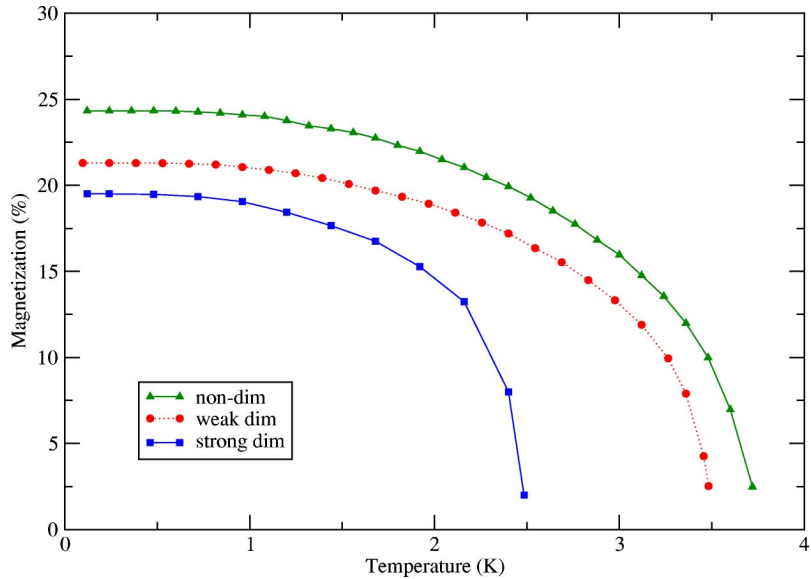

FIG. 2. Representative order parameter curves for the dimerized and nondimerized chains obtained numerically. The triangles represent data for the nondimerized chain, circles for weak dimerization, and squares for strong dimerization. Dimerization is seen to reduce both the Néel temperature $T_N$ and the zero temperature order parameter $m_c$. For the nondimerized chain the initial intrachain coupling distribution is taken to be uniform with the couplings lying in the range [0,1]; for the dimerized chain with weaker dimerization even bonds are uniformly distributed between [0,0.9] and odd bonds between [0.1,1], while for the dimerized chain with stronger dimerization even bonds are uniformly distributed between [0,0.8] and odd bonds between [0.2,1]. The interchain coupling for the shown results was taken to be $zJ_\perp/J_0=0.2$, where $J_0$ is the cutoff for the initial intrachain coupling distribution. The scale for the temperature axis is obtained by setting $J_0=120$ K: the value for the intrachain coupling realized in CuGeO$_3$.

Using an interchain coupling of $zJ_\perp/J_0=0.2$, where $J_0$ is the cutoff for the initial intrachain coupling distribution, and initial distributions corresponding to random dimerized and nondimerized chain, we solve for the order parameter as a function of the temperature. The randomness is determined by the width of the initial distribution of couplings while the dimerization depends on the offset between the even and odd coupling distributions. A representative set of order parameter curves obtained for the nondimerized and dimerized chains with different dimerization strength are shown in Fig. 2. For the nondimerized chain the initial intrachain coupling distribution is taken to be uniform with the couplings lying in the range [0,1]; for the dimerized chain with weaker dimerization even bonds are uniformly distributed between [0,0.9] and odd bonds between [0.1,1], while for the dimerized chain with stronger dimerization even bonds are uniformly distributed between [0,0.8] and odd bonds between [0.2,1]. The choice of the interchain coupling and the scale for the temperature with $J_0\sim 120$ K are motivated by the parameters realized in CuGeO$_3$. It is clear that increasing dimerization strength suppresses both $m_c$ and $T_N$, which is what we expect and also consistent with experiments. However, within the approach used here it is not possible to make a quantitative comparison of the results with experiments since the distribution of couplings realized in doped CuGeO$_3$ is not known.

We have also verified that increasing disorder strength enhances both $m_c$ and $T_N$, for fixed interchain coupling. Specifically, we study nondimerized chains with the initial distribution taken to be the uniform box distribution centered at $J_m=0.5$, with width $w$; i.e., $P(J)=1/w$ for $J_m-w/2<J<J_m+w/2$ and $P(J)=0$ otherwise. The renormalized width of the distribution $w^*=w/J_m$ is a measure of the degree of randomness. The interchain coupling in this case, expressed in terms of the mean coupling $J_m$, is taken to be $zJ_\perp/J_m=0.4$. The resulting order parameter curves for different $w^*$ values are shown in Fig. 3. With increasing randomness, the proportionality constant between $T_N$ and $m_c$ moves closer to its analytically predicted value $zJ_\perp/2$ (see inset of Fig. 3). Due to the large interchain coupling, the system is far from the weak coupling limit; hence, as expected, $T_N/m_c$ is not close to $zJ_\perp/2$ (0.1).

The numerical approach can also be used to study systems with weaker interchain coupling. For lower values of interchain coupling, the numerically obtained renormalized order parameter curve approaches the analytically obtained universal curve (Fig. 4). Further, as shown in the inset of Fig. 4, for lower interchain coupling, the renormalized proportionality constant c* $(2T_N/zJ_\perp m_c)$ gets closer to 1.0, its analytically predicted value. The universal term in magnetization [Eq. (7)] is dominant over other contributions only as a logarithm; thus the universal ratio is approached logarithmically as interchain coupling strength goes to zero. As discussed above, the increase in randomness also leads to a similar, though less pronounced trend. In both cases the system is closer to the random singlet phase which is the basis for the analytical results.

Consistent with the results in the weak coupling limit, disorder leads to, or enhances the tendency toward AF LRO (Fig. 3). Dimerization in the random chain is seen to reduce both the Néel temperature $T_N$ and the zero temperature order parameter $m_c$ (Fig. 2). In the dimerized chain, the Néel temperature shows a sharp reduction with decreasing randomness or increasing dimerization, and the AF LRO vanishes in the weak randomness/strong dimerization limit. The effect of randomness is less pronounced in the nondimerized chain with both $T_N$ and $m_c$ being less sensitive to the amount of randomness. These results are consistent with experiments in quasi-one-dimensional materials.[4,7,10,20] In particular, the disorder induced ordering, and the behavior of the Néel temperature and order parameter in presence of randomness are in good qualitative agreement with the observations in the spin-Peierls material $CuGeO_3$.

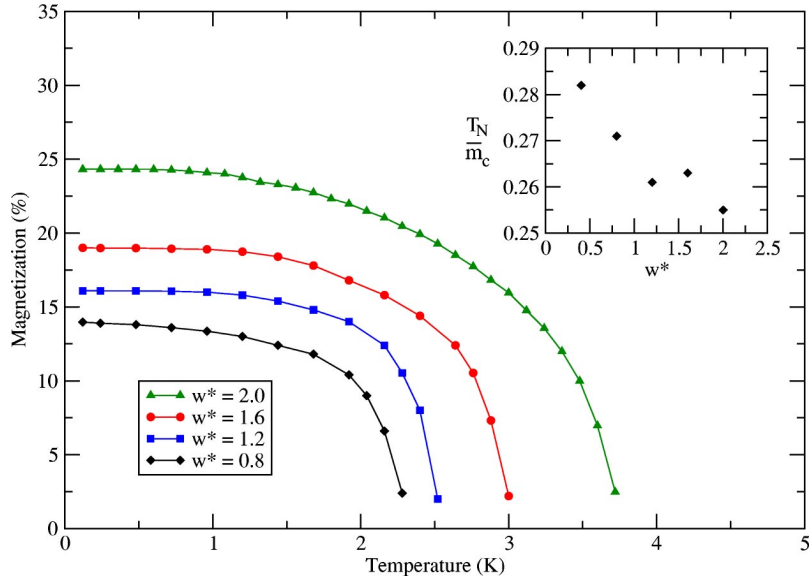

FIG. 3. Representative order parameter curves for the nondimerized chain showing the effect of randomness on $T_N$ and the order parameter. Increase in randomness enhances the tendency towards AF long-range order, resulting in larger $T_N$ and zero temperature order parameter $m_c$. The initial intrachain coupling distribution is taken to be the uniform box distribution centered at $J_m = 0.5$, with width $w$; i.e., $P(J) = 1/w$ for $J_m - w/2 < J < J_m + w/2$ and $P(J) = 0$ otherwise. The renormalized width of the distribution $w^* = w/J_m$ is a measure of the degree of randomness. The interchain coupling for the shown results was taken to be $zJ_\perp/J_m = 0.4$. For consistency with results shown in Fig. 2, the scale for the temperature axis is obtained by setting $J_m = 60$ K. The inset shows the proportionality constant $T_N/m_c$ as a function of degree of randomness ($w^*$). With increasing randomness, the ratio $T_N/m_c$ tends towards its analytically predicted value $zJ_\perp/2$ (0.1). Due to the large interchain coupling, the system is far from the weak coupling limit; hence, $T_N/m_c$ is, as expected, not close to 0.1.

In our work we have modeled the effect of doping by randomness in the coupling between neighboring spins. In the case of $CuGeO_3$, this corresponds more closely to Si doping, because the Si goes to the Ge sites, without affecting the half spins carried by the Cu ions; their main effect is changing the coupling strength by distorting the local lattice structure. The situation is quite different for Zn, Ni, or Mg doping, as they go directly to the Cu sites and either remove or change the size of the spins of the sites; they are thus *not* properly described by our model. The mean-field treatment of interchain coupling becomes exact in the limit $z \to \infty$; in real systems such as $CuGeO_3$, however, $z$ is finite, and thus there are both quantum and spatial fluctuations of the local field $h$. Such fluctuations which are not included in our study can become very important in the ultimate strong randomness limit, and are likely to be the reason why Néel order is suppressed for sufficiently large doping concentration.

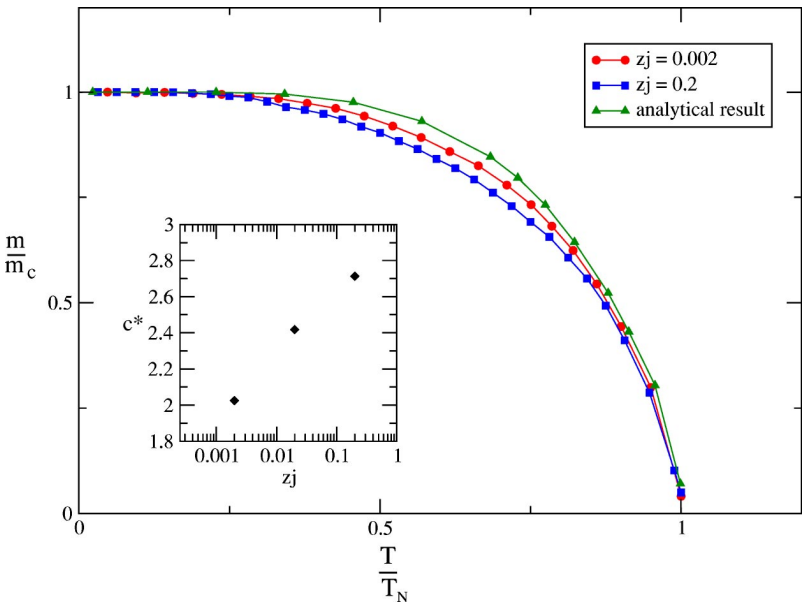

FIG. 4. Renormalized order parameter curves for the nondimerized chain with varying interchain coupling. As the interchain coupling is lowered, the order parameter curve tends toward the analytically obtained universal curve. The triangles represent the universal curve, while squares and circles represent the numerical result for $zJ_\perp/J_0 = 0.2$ and $zJ_\perp/J_0 = 0.002$, respectively ($J_0$ is the cutoff for the initial intrachain coupling distribution). The numerical curves are obtained using a uniform initial intrachain distribution of couplings lying in the range [0,1]. The inset shows the renormalized proportionality constant $c^*$ ($2T_N/zJ_\perp m_c$) as a funtion of interchain coupling $zj$ ($zJ_\perp/J_0$). The constant $c^*$ gets closer to 1.0, its analytically predicted value, as the interchain coupling is lowered.

The issue of disorder driven LRO in materials such as $CuGeO_3$ has been addressed theoretically by a number of authors.[23–34] Most of these studies model Zn, Ni, or Mg doping,[24,25,27–30,32,33] namely, some of the half spins are either removed (Zn or Mg doping), or replaced by $S=1$ spins (Ni doping), while the case of Si doping has been addressed by some workers[23,26,31,34] using models similar to that used in the present work. Reference 34 presented results of finite-size quantum Monte Carlo studies, in which they modeled the effect of Si doping as bond dilution. Their conclusion that a finite concentration of dopant is needed to stabilize a Néel ordered ground state is in agreement with our results that finite randomness is needed to induce Néel order in the presence of dimerization. References 23,26,31 use the mean-field treatment of interchain coupling similar to what is done here, as well as in other works on Zn, Ni, or Mg doping.[25,27,28,33] On the other hand, their treatment of the resultant intrachain model is very different from ours. Reference 23 treated the intrachain spin correlation at the semiclassical level which leaves out quantum fluctuations, while our RSRG treatment of the intrachain physics is fully quantum mechanical, and asymptotically exact. Reference 31 used an $XY$ model for intrachain coupling, even though the Néel order parameter is assumed to be along the $\hat{z}$ direction; the reason that the $XY$ model is used is because it can be mapped onto free fermions using the Jordan-Wigner transformation. Reference 26 starts with the Heisenberg model but uses Hartree-Fock approximation to treat the $S_z \cdot S_z$ coupling, thus reducing it to an $XY$ model. We believe the RSRG treatment of the effective 1D problem in the present work has the following advantages. (i) The fact that the RSRG approach is fully quantum mechanical and asymptotically exact in the low-energy limit ensures that our results, including the leading dependence of the Néel order parameter ($m$) and $T_N$ on the interchain coupling strength, and the $T$ dependence of $m$, are exact in the weak interchain coupling limit within the mean-field treatment of interchain coupling; since the latter is exact in the limit of large dimensionality or coordination number, our results become exact in that limit with weak interchain coupling. (ii) This approach preserves the rotational symmetry of the original spin Hamiltonian, which is known to be respected in $CuGeO_3$ to a high degree; therefore, the approach is conceptually closer to the situation realized in $CuGeO_3$. (iii) The Jordan-Wigner transformation is specific to half spins, while our approach can be easily generalized to study other types of coupled spin chains, including spins of bigger size, or in the presence of further neighbor couplings along the chain. We briefly discuss an example of the former. RSRG study of AF spin-1 chain[22] demonstrated a second-order phase transition from the Haldane phase to the random singlet phase as bond randomness increases; there is a Griffith region in the Haldane phase close to the critical point, where the spin susceptibility is divergent. We thus expect with *weak* interchain coupling the system becomes Néel ordered at $T=0$, when the single chain is either in the random singlet phase or in the Griffith region, with the low-$T$ behavior following that of the weakly coupled undimerized and dimerized spin-1/2 chains described here, respectively.[35,36] We also note that in addition to the differences in the approaches used, our work further differs from previous works in that we focus on physical quantities that have not been previously calculated in this particular case, namely, the temperature dependence of the magnetization.[37] Previous works,[23,26,31] on the other hand, have been mostly focused on the onset of Néel ordering, and the dependence of the Néel temperature on interchain coupling strength and doping level (which we also study here). The temperature dependence of intensity of the AF peak has been studied experimentally for cases of both Zn and Si doping.[38–40] It is possible to obtain the magnetization from the intensity of the AF peak.[40] If the experimental measurements for the AF peak intensity are mapped to obtain the temperature dependence of magnetization, a direct comparison between our results and experiments will be possible and would certainly be of interest.

In conclusion, using a combination of mean-field treatment of interchain coupling and real space renormalization group treatment of the resultant single chain spin Hamiltonian, we demonstrated that bond randomness in the single chain Hamiltonian enhances the tendency toward Néel ordering. Our results compare favorably with the phenomenology of doped $CuGeO_3$.

This work was supported in parts by NSF Grants No. DMR-9971541 and No. DMR-0225698 (K.Y.), the state of Florida (A.J.), and the A. P. Sloan Foundation (K.Y.). A.J. would like to thank E. Yusuf and J. R. Schrieffer for useful discussions and comments.